\documentclass{article}
\usepackage{ijcai19}
\usepackage{multirow}
\usepackage{tabu}
\usepackage{times}
\usepackage{soul}
\usepackage{url}
\usepackage[draft]{hyperref}
\usepackage[utf8]{inputenc}
\usepackage[small]{caption}
\usepackage{graphicx}
\usepackage{epsfig}
\usepackage{amsmath}
\usepackage{booktabs}
\usepackage{amsfonts,amssymb}
\usepackage{color}
\newcommand{\nop}[1]{}

\title{Convolutional Gaussian Embeddings for\\Personalized Recommendation with Uncertainty}

\author{
Junyang Jiang$^1$\and
Deqing Yang$^{2}$
\footnote{Contact Author}
\and
Yanghua Xiao$^{1,3}$\And
Chenlu Shen$^2$\\
\affiliations
$^1$School of Computer Science, Shanghai Key Laboratory of Data Science, Fudan University, China\\
$^2$School of Data Science, Fudan University, China\\
$^3$Shanghai Institute of Intelligent Electronics \& Systems, China\\
\emails
{
\{jiangjy15, yangdeqing, shawyh, clshen17\}@fudan.edu.cn}
}

\begin{document}

\maketitle

\begin{abstract}
Most of existing embedding based recommendation models use embeddings (vectors) corresponding to a single fixed point in low-dimensional space, to represent users and items. Such embeddings fail to precisely represent the users/items with uncertainty often observed in recommender systems. Addressing this problem, we propose a unified deep recommendation framework employing Gaussian embeddings, which are proven adaptive to uncertain preferences exhibited by some users, resulting in better user representations and recommendation performance. Furthermore, our framework adopts Monte-Carlo sampling and convolutional neural networks to compute the correlation between the objective user and the candidate item, based on which precise recommendations are achieved. Our extensive experiments on two benchmark datasets not only justify that our proposed Gaussian embeddings capture the uncertainty of users very well, but also demonstrate its superior performance over the state-of-the-art recommendation models.
\end{abstract}
\vspace{-0.2cm}

\section{Introduction}
Recommender systems have demonstrated great commercial value in the era of information overload, because they help users filter our their favorite items precisely from large repositories. No matter in traditional matrix factorization (MF for short) based models \cite{NMF,SVD++} or in recent deep neural models \cite{NCF,NFM}, users and items are generally represented as low-dimensional vectors, also known as \emph{embeddings}, which are learned from observed user-item interactions or user/item features. In these models, a user/item representation is a single fixed point of the continuous vector space, which represents a user's preferences or an item's characteristics. Then, the final recommendation results are generated through computing the correlations between user embeddings and item embeddings, such as inner product of two embeddings \cite{NAIS} or feeding them into multi-layer perceptron (MLP for short) \cite{DRM,NCF}.

\begin{figure}[b]
\vspace{-0.1cm}
\center
\epsfig{file=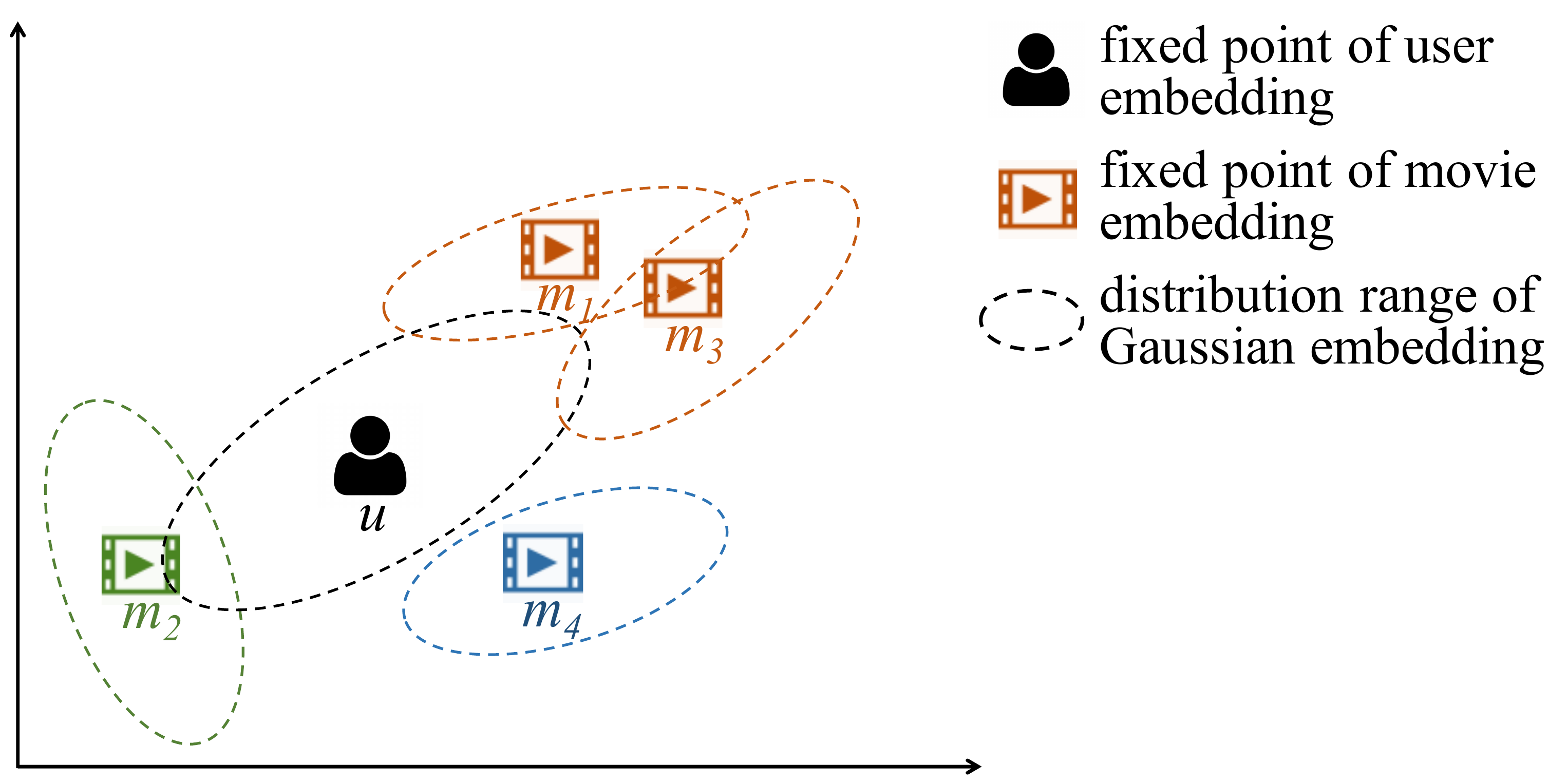,width=2.5in}
\vspace{-0.2cm}
\caption{The fixed points of user/item embeddings can not well represent the users/items with uncertainty, resulting in inaccurate recommendation. While distribution based embeddings handle the ones with uncertainty well.}\label{fig:example}
%\vspace{-0.3cm}
\end{figure}

Despite their successes, one unneglectable limitation of these embedding-based models is the lack of handling uncertainty. In a recommender system, users may induce uncertainty due to some reasons. One reason is the lack of discriminative information \cite{zhu2018deep}, especially for those users who have very few or even no observed user-item interactions, e.g., historical ratings or reviews for items. Even for the users who have sufficient interactions, uncertainty may also be caused by diversity \cite{bojchevski2018deep}, e.g., some users exhibit many and very distinct genres of preferences. We illustrate the example in Figure \ref{fig:example} to explain why the embeddings corresponding to fixed points can not well handle such cases. Suppose user $u$ has rated movie $m_1$ and $m_2$ with high scores and these two movies belong to very distinct genres which are labeled with different colors. If we use fixed embeddings learned from observed user-movie interactions to represent users and movies, $u$'s position in the embedding space (mapped into a 2D map) may locate in the middle of $m_1$ and $m_2$. If the recommendation is made based on the distance between the embedding positions of $u$ and the candidate movies, $u$ may be recommended to with movie $m_4$ of the genre different from $m_1$ and $m_2$, instead of $m_3$ of the same genre as $m_2$, because $u$ is closer to $m_4$ than $m_3$. There is another case that $u$'s position may be closer to $m_2$ than to $m_1$, then $m_3$ still has fewer chances to be recommended to $u$. %For those users with uncertainty, the representations of single fixed points fail to capture their preferences very well, resulting in limited recommendation performance.

In recent years, some researchers have employed \emph{Gaussian embeddings} to learn the representations of words \cite{vilnis2014word} and nodes in a graph \cite{bojchevski2018deep,zhu2018deep} because of their advantage on capturing the uncertainty in representations. It motivates us to employ Gaussian embeddings in recommender systems to represent users and items more flexibly. In Gaussian embedding based mechanism, each user or item embedding corresponds to a Gaussian distribution of which the mean and the variance are learned from observed user-item interactions. In other words, each user or item is represented by a density instead of a fixed point in latent feature space. The variance of the Gaussian distribution of a user/item measures uncertainty associated to the user/item's representation. Recall the example in Figure \ref{fig:example}, if $u$ and all movies are represented by Gaussian embeddings, their positions in the space are the distribution ranges labeled by the dashed ellipses rather than fixed points. As depicted in the figure, $u$'s range may overlap $m_3$'s range other than $m_4$'s range. Thus precise recommendation results for the users with uncertainty are achieved.

%In this paper, we design a deep recommendation framework with Gaussian embeddings to accomplish the recommendation task of implicit feeadback \cite{NCF,NAIS}. 
Most of existing Gaussian embeddings based models are learned based on ranking-based loss \cite{dos2017gaussian,vilnis2014word}, which is not applicable to the tasks other than learning to rank, such as predicting a value or classification. This is because metrics used in previous work, such as KL-divergence, take on a more limited range of values, which is not enough for the input of a classifier \cite{vilnis2014word}. Besides, models for rate prediction rely on an absolute, rather than relative manner. Therefore it is not feasible to employ such a ranking scheme for the recommendation tasks other than ranking candidate items. 

In this paper, we propose a recommendation framework in terms of implicit feedback \cite{NCF,NAIS} rather than only ranking candidate items. As a result, we adopt a learning principle different from previous ranking-based Gaussian embedding models. Specifically, our framework first learns a Gaussian distribution for each user and item. Then, according to the distribution, a group of samples is generated through \emph{Monte-Carlo} sampling \cite{MC} for the objective user and the candidate item, respectively. The generated samples are used to compute the correlation between the user and the item, based on which precise recommendation results are achieved. Furthermore, in order to compute the correlation between the user and the item effectively, our framework incorporates convolutional neural network (CNN for short) to extract and compress the features from the user-item sample pair. Our experiment results have proven such convolutional operation is more effective than previous average-based method \cite{joon2019iclr}.

In our framework, if the user and the item are regarded as two objects, the correlation computed based on their Gaussian embeddings actually quantifies the matching degree of the two objects. Therefore, our framework can be extended to other machine learning tasks such as link prediction and classification. 

In summary, the contributions of our work include:

1.  We employ Gaussian embeddings into recommender systems to represent users and items, which is more effective than traditional embeddings of fixed points.

2. We adopt convolutional operations to learn the correlation between the Gaussian embeddings of an objective user and a candidate item efficiently, which is critical to generating precise recommendation results.

3. The extensive experiments conducted on two benchmark datasets justify that our proposed Gaussian embeddings capture the uncertainty of some users well, and demonstrate our framework's superiority over the state-of-the-art recommendation models.

%However, a point vector can not model uncertainty. For example, given a user is interested in both comedy drama and animation. After training with previous point vector embedding-based recommendation algorithms, the user's embedding is close to the group of comedy drama people or animation people. 

The rest of this paper is organized as follows. We present the design details of our framework in Section 2, and show our experimental results in Section 3. In Section 4, we introduce related work and conclude our work in Section 5. 

%\vspace{-0.1cm}
\section{Methodology}
\subsection{Problem Statement}
%We firstly formalize the problem addressed in this paper, followed by the introduction of the preliminaries of our work. 
In the following introduction, we use a bold uppercase to represent a matrix or a cube, and a bold lowercase to represent a vector unless otherwise specified,.
%\vspace{-0.1cm}
\subsubsection{Implicit Feedback} We design our framework in terms of implicit feedback which is also focused in many recommendation models \cite{CKE,NCF,NAIS}. Given a user $u$ and an item $v$, we define observed $u$'s implicit feedback to $v$ as
\vspace{-0.2cm}
$$
\vspace{-0.1cm}
y_{uv}=
\begin{cases}
 1 & \text{if $u$ interacts with $v$, such as rating or review}\\
 0 & \text{otherwise}
\end{cases}
$$

The task of our framework is predicting a given objective user $u$'s implicit feedback to a candidate item $v$, which is essentially a binary classification model. Accordingly, our framework should estimate the probability that $u$'s implicit feedback to $v$ is 1, which is denoted as $\hat{y}_{uv}$ in this paper.
%\vspace{-0.1cm}
\subsubsection{Gaussian Embedding}
%In our framework, each user or item is represented as a lower-dimensional Gaussian distribution vector (embedding) $\boldsymbol{h}=\mathcal N (\boldsymbol{\mu}, \boldsymbol{\Sigma})$ where $\boldsymbol{\mu}\in\mathbb{R}^D, \boldsymbol{\Sigma}\in\mathbb{R}^{D\times D}$, and $D$ is embedding dimension. Such latent representations should preserve the similarities between users and items in the embedding space, based on which the correlations between users and items are evaluated. As a result, given a user $u$ and an item $v$, our framework tries to evaluate $\hat{y}_{uv}$ based on $g_u$ and $g_v$.

In our framework, each user or item is represented by a Gaussian distribution consisting of a expectation embedding (vector) and a covariance matrix, i.e., $g=\mathcal N (\boldsymbol{\mu}, \boldsymbol{\Sigma})$ where $\boldsymbol{\mu}\in\mathbb{R}^D, \boldsymbol{\Sigma}\in\mathbb{R}^{D\times D}$, and $D$ is embedding dimension. Such latent representations should preserve the similarities between users and items in the embedding space, based on which the correlations between users and items are evaluated. As a result, given a user $u$ and an item $v$, our framework tries to evaluate $\hat{y}_{uv}$ based on $g_u$ and $g_v$.

More specifically, to limit the complexity of the model and reduce the computational overhead, we assume that the embedding dimensions are uncorrelated \cite{bojchevski2018deep}. Thus $\boldsymbol{\Sigma}$ is considered as a diagonal covariance matrix $diag(\Sigma_1, \Sigma_2, \cdots, \Sigma_D)$ and can be further represented by a $D$-dimensional array.

%\vspace{-0.1cm}
\subsubsection{Algorithm Objective}\label{sec:alg}
Before describing the details of our proposed framework, we first summarize our algorithm's objective. Formally, we use $p(l|u,v)$ to denote the probability that the matching degree between user $u$ and item $v$ is $l$. In our scenario of implicit feedback, $l$ is either 1 or 0, and we denote $p(l=1|u,v)$ as $\hat{y}_{uv}$. Therefore, the $p(l=1|u,v)$ of high value indicates that we should recommend $v$ to $u$. If $l$ is labeled with a rating score, $p$ can be used to predict $u$'s rating score on $v$, which reflects the degree of $u$'s preference to $v$. Moreover, $p(l=1|u,v)$ can be used to indicate a classification task if $l$ is regarded as class label.  

According to the aforementioned problem definition, $p(l|u,v)$ is estimated with $p(l|g_u, g_v)$. Recall that Gaussian distribution is a probability distribution of a random variable, so we calculate $p(l|g_u, g_v)$ as
\begin{equation}\label{eq:p1}
\vspace{-0.1cm}
p(l|g_u, g_v)=\int p(l|\boldsymbol{z}_u,\boldsymbol{z}_v)p(\boldsymbol{z}_u|g_u)p(\boldsymbol{z}_v|g_v)d\boldsymbol{z}_u d\boldsymbol{z}_v
%\vspace{-0.1cm}
\end{equation}
where $\boldsymbol{z}_u, \boldsymbol{z}_v \in \mathbb{R}^D$ are the vectors of random variables sampled based on Gaussian distribution $g_u$ and $g_v$, respectively. 

To approximate the integration in Eq.\ref{eq:p1}, we adopt Monte-Carlo sampling \cite{MC}. Specifically, suppose that we sample $\boldsymbol{z}_u \sim g_u$ and $\boldsymbol{z}_v
\sim g_v$ for $K$ times, then we have
\vspace{-0.1cm}
\begin{equation}\label{eq:p2}
p(l|g_u,g_v) = \lim_{K\to \infty} \frac{1}{K^2}\sum_{i=1}^{K}\sum_{j=1}^{K}p(l|\boldsymbol{z}_u^{i}, \boldsymbol{z}_v^{j})
\end{equation}

The calculation of Eq.\ref{eq:p2} is challenging. On one hand, a large $K$ incurs unbearable computational cost. On the other hand, a small $K$ incurs bias, resulting in unsatisfactory recommendation results. What is more, it is not trivial to compute $p(l|\boldsymbol{z}_u^i,\boldsymbol{z}_v^j)$. In fact, we can rewrite Eq.\ref{eq:p2} as
\begin{equation}\label{eq:p3}
p(l|g_u,g_v) =p\bigg(l\bigg|
\left\{
  \begin{array}{ccc}
          (\boldsymbol{z}_u^1,\boldsymbol{z}_v^1) & \cdots & (\boldsymbol{z}_u^1,\boldsymbol{z}_v^K)\\
          \vdots & \ddots & \vdots\\
          (\boldsymbol{z}_u^K,\boldsymbol{z}_v^1) & \cdots & (\boldsymbol{z}_u^K,\boldsymbol{z}_v^K)
  \end{array}
  \right\}
  \bigg)
\end{equation}This formula implies that $p$ is computed based on $K^2$ correlations of vector pair $(\boldsymbol{z}_u^{i},\boldsymbol{z}_v^{j})$. Inspired by CNN's power on extracting and compressing features in image processing, we choose a CNN fed with the $K^2$ vector pairs to compute Eq.\ref{eq:p3}, in which the convolution kernels are used to learn the pairwise correlations of  $(\boldsymbol{z}_u^{i},\boldsymbol{z}_v^{j})$. The computation details will be introduced in the next subsection. Our experiment results will prove that the CNN-based computation of Eq.\ref{eq:p3} is more effective than computing the mean of $p(l|\boldsymbol{z}_u^{i}, \boldsymbol{z}_v^{j})$.

%Inspired by [], we choose to use neural Networks to learn the Equation (3) with the help of its generalization capacity and expressive power. In comparison with simple element-wise operation like averaging or euclidean distance, deep neural networks can automatically model more complex relationships. And generalization ability of neural networks can preventing bulk sampling. More specifically, we apply a convolutional neural network (CNN) because we can rewrite Equation (3) as:

%More generally, to learn proximity between user and item represented by two Gaussian distribution , we have the option of applying Neural Networks to learn from interactions between vectors sampled by these Gaussian distribution.

\subsection{Framework Description}
%Next, we introduce the details of our recommendation framework which is designed to compute Eq.\ref{eq:p3} effectively. As depicted in Figure \ref{fig:frame}, our framework consists of four layers which are described in turn from initial input to final output in this subsection.

\begin{figure}[!htb]
%\vspace{-0.2cm}
\center
\epsfig{file=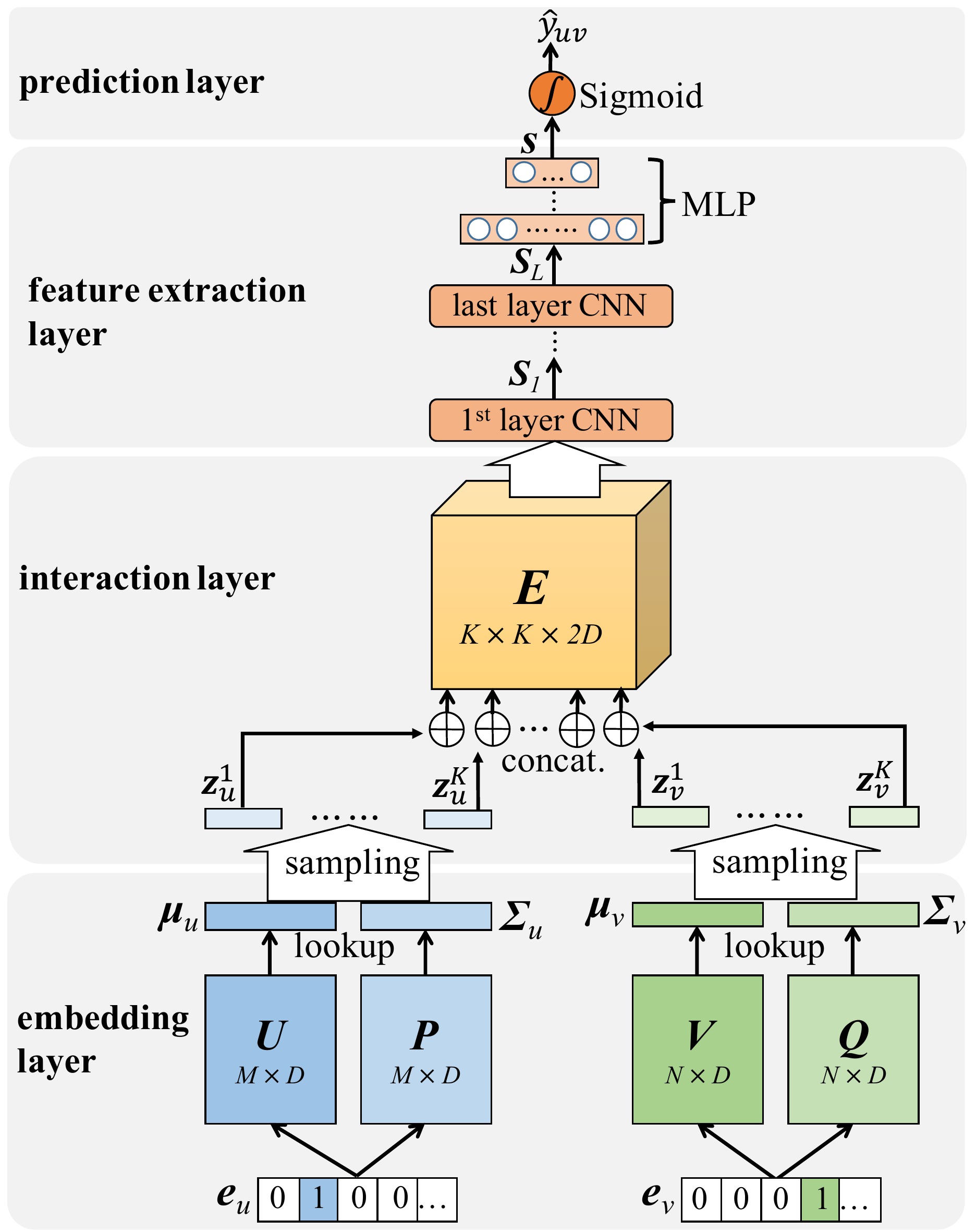,width=3.15in}
\vspace{-0.2cm}
\caption{The overview of our recommendation framework.}\label{fig:frame}
\vspace{-0.2cm}
\end{figure}

\subsubsection{Embedding Layer}
The first layer is the \emph{embedding layer}. At first, a user $u$ and an item $v$ are represented as a one-hot vector of $D$ dimensions, denoted as $\boldsymbol{e}_u\in\mathbb{R}^D$ and $\boldsymbol{e}_v\in\mathbb{R}^D$, respectively. Besides $u/v$'s ID, the dimensions of value 1 in $\boldsymbol{e}_{u/v}$ can also correspond to $u/v$'s feature IDs. In our experiments, we only input user/item IDs into our framework. Furthermore, we initialize four embedding matrices $\boldsymbol{U},\boldsymbol{P}\in \mathbb{R}^{M\times D}$ and $\boldsymbol{V},\boldsymbol{Q}\in \mathbb{R}^{N\times D}$ where $M$ and $N$ are user (or user feature) number and item (or item feature) number, respectively. Then, we have the Gaussian mean vector and variance vector of $u$ and $v$ through the following lookup operation,
\begin{align}
\vspace{-0.1cm}
\boldsymbol{\mu}_u = \boldsymbol{U}^T\boldsymbol{e}_u, \boldsymbol{\Sigma}_u= ELU(\boldsymbol{P}^T\boldsymbol{e}_u)+\boldsymbol{1}\\
 \boldsymbol{\mu}_v= \boldsymbol{V}^T\boldsymbol{e}_v,
\boldsymbol{\Sigma}_v = ELU(\boldsymbol{Q}^T\boldsymbol{e}_v)+\boldsymbol{1}
\end{align}
where $\boldsymbol{1}\in\mathbb{R}^D$ is an array filled with value 1 and $ELU$ is Exponential Linear Unit. Both of them are used to guarantee that every element of variance vector is non-negative. Thus, we get $g_u =\mathcal N (\boldsymbol{\mu}_u, \boldsymbol{\Sigma}_u)$ and $g_v =\mathcal N (\boldsymbol{\mu}_v, \boldsymbol{\Sigma}_v)$.

%Following [], in our experiments, the feature vectors are only user and item ID feature and thus M and N are the number of users and number of items respectively.
\subsubsection{Interaction Layer}
The second layer is the \emph{interaction layer}, in which $K$ samples are sampled for $u$ and $v$ according to the Monte-Carlo sampling under $g_{u}$ and $g_v$, respectively. In order to perform backpropagation, we use the reparameterization trick in \cite{kingma2013auto} to obtain the embedding of $u$'s $i$-th sample as follows
\begin{align}
\vspace{-0.2cm}
\boldsymbol{z}_u^i = \boldsymbol{\mu}_u + \boldsymbol{\Sigma}_u^{\frac{1}{2}} \times \epsilon
\end{align}
where $\epsilon$ is an auxiliary noise variable $\epsilon \sim \mathcal{N}(0, 1)$ and varies in each sample. So does $\boldsymbol{z}_v$.

As stated in subsection \ref{sec:alg}, $p(l|g_u,g_v)$ is computed based on the correlations of $K^2$ sample pairs. Hence in this layer we construct a big \emph{interaction map} $\boldsymbol{E}$ consisting of $K^2$ units. Each unit $\boldsymbol{E}_{(i,j)}$ represents the correlation of a sample pair $(\boldsymbol{z}_u^i,\boldsymbol{z}_v^j)$, which has the following expression, 
\begin{equation}
\vspace{-0.1cm}
\boldsymbol{E}_{(i,j)}=[\boldsymbol{z}_u^{i}, \boldsymbol{z}_v^{j}]
\end{equation}
where $[\cdot, \cdot]$ is the concatenation of two vectors. As a result, $\boldsymbol{E}$ is actually a cube of $K \times K \times {2D}$ dimension. Then, we should utilize $\boldsymbol{E}$ to compute $p(l|g_u,g_v)$, which is implemented in the next layer.%In such case, every vector sampled from user have a chance to interact with all vectors sampled from item.

We note that other interaction operations of two vectors, such as inner product and element-wise product, are also widely used in other recommendation models. But our empirical results show that concatenation outperforms other functions consistently. One possible explanation is concatenation preserves original feature of two vectors and thus neural networks can better learn their proximity. %A more principled way is to combine the more than one interaction function and jointly train the framework, which we leave as future work.

\subsubsection{Feature Extraction Layer}
The input of this layer is the output of the preceding interaction layer, i.e., the cube $\boldsymbol{E}$. In fact, $\boldsymbol{E}$ contains rich features that are beneficial to compute Eq.\ref{eq:p3}. It is analogous to an image containing pixel features except that the number of channels is $2D$. Inspired by the usage of CNN for extracting and compressing object features which has been proven effective in the field of computer image processing \cite{krizhevsky2012imagenet}, we employ a multi-layer CNN followed by an MLP in this feature extraction layer. 
%%%%%%%%%%%%%%%%%%%
\nop{
\begin{align}
    \boldsymbol{s}_1 &= \phi_1(\boldsymbol{O}_1*\boldsymbol{E})\\
    \boldsymbol{s}_2 &= \phi_2(\boldsymbol{O}_2*\boldsymbol{s}_2)\\
    \cdots\\
    \boldsymbol{s}_L &= \phi_L(\boldsymbol{O}_L*\boldsymbol{s}_{L-1})
\end{align}
where $\boldsymbol{O}$ denotes the operation in neural network like convolution, pooling and fully-connected layer; $\phi$ denotes activation function like ReLU. To meet the theoretical motivation described in Section 2.2 perfectly, we employ convolution layer as $\boldsymbol{O}_1$. Besides, there is another reason for using CNN is that CNN can reduce the dimensions with fewer parameters. In contrast, if we simply use MLP, there would be large parameters especially when we have a large $D$, which would require more training computation and cause for inefficiency.    
}%%%%%%%%%%%%%%%%
%In our experiments, following the popular and fundamental fashion of designing networks for image classification\cite{krizhevsky2012imagenet}, we apply a simple CNN structure with 2 convolutional layers followed by MLP. 

Specifically, for each layer of the CNN, we apply $T$ filters $\boldsymbol{G}\in \mathbb{R}^{l_k\times l_k \times c}$ to extract specific local patterns where $l_k\times l_k $ is kernel (window) size of the filter, and $c$ is channel number. We feed $\boldsymbol{E}$ into the first layer of our CNN to generate its output $\boldsymbol{S}_1$ as follows
\begin{equation}
%\vspace{-0.1cm}
 \boldsymbol{S}_1= ReLU(\boldsymbol{G}_1 \otimes \boldsymbol{E}+b_1)
\end{equation}
where $\boldsymbol{G}_1\in \mathbb{R}^{l_k\times l_k \times 2D}$, $\otimes$ is convolution operator and $b_1$ is bias. In general, $T$ is set to a large number such as 32 or 64, which helps us learn more than one correlation of each vector pair. Besides, one filter computes the correlation of exactly one vector pair if we set $l_k$=1. Otherwise, the filter extracts features from adjacent vector pairs. In different layers of the CNN, we can set different $l_k$s. Our empirical results show that a larger kernel size greatly reduces computing cost but contributes little to overall accuracy. Another reason for adopting convolution is that it can reduce the dimensions with fewer parameters. 

For each of the rest layers of the CNN, its input is the output of the preceding layer. Suppose $\boldsymbol{S}_L$ is the output of the CNN's last layer, all of $\boldsymbol{S}_L$'s features are flattened through concatenation to be fed into an MLP to obtain final output of this feature extraction layer, i.e.,
\begin{equation}
\boldsymbol{s} = MLP([\boldsymbol{S}_L^1\, \boldsymbol{S}_L^2\, \cdots \,\boldsymbol{S}_L^T])
\end{equation}
where $\boldsymbol{s}\in \mathbb{R}^{D'}$ and $\boldsymbol{S}_L^i (1\leq i\leq T)$ is the flattened array of feature map corresponding to the $i$-th filter. In the following evaluation of our framework, we adopted a CNN of two layers, i.e., $L$=2. The first layer's $l_k$ is set to 1, and the second layer's $l_k$ is set to 2.  

\subsubsection{Prediction Layer}
The last layer of our framework is the prediction layer, which accepts the output of the preceding CNN, i.e., $\boldsymbol{s}$, to generate the final prediction score $\hat{y}_{uv}$. In this layer, we feed  $\boldsymbol{s}$ into a single layer perceptron and use Sigmoid function $\sigma(\cdot)$ to compute $\hat{y}_{uv}$ as follows
\begin{equation}
    \hat{y}_{uv}=\sigma(\boldsymbol{W}_{out}^T\boldsymbol{s}+b)
\end{equation}
where $\boldsymbol{W}_{out}^T\in \mathbb{R}^{D'}$ is the weight matrix and $b$ is a bias vector. According to $\hat{y}_{uv}$, we can decide whether $v$ deserves being recommended to $u$.

\subsubsection{Model Learning}
To learn our model's parameters including all embeddings mentioned before, we use \textit{binary cross-entropy loss} since it is suitable for binary classification. Specifically, we have
\begin{equation}\label{eq:loss}
    \mathcal L = -\bigg\{\sum_{(u,v)\in \mathcal Y^+ \cup \mathcal Y^-}y_{uv}\log \hat{y}_{uv}+(1-y_{uv})\log(1-\hat{y}_{uv})\bigg\}
\end{equation}
where $\mathcal Y^+$ denotes the set of observed interactions ($\hat{y}_{uv}=1$), and $\mathcal Y^-$ denotes the set of negative instances which are sampled randomly from unobserved interactions. In our experiments, we use Adam algorithm \cite{Adam}
%where learning rate is 1e-3, $\beta$s are 0.9 and 0.999 and $\epsilon$ is 1e-8, 
to optimize Eq.\ref{eq:loss}, because it has been proven to be powerful optimization algorithm for stochastic gradient descent for training deep learning models.

Please note that our framework can be applied to various recommendation tasks, including personalized ranking and rating prediction, through simply modifying the loss function.

%\vspace{-0.1cm}
\section{Experiments}
In this section, we conduct extensive experiments to answer the following research questions.

\emph{RQ1}: Which hyper-parameters are critical and sensitive to our framework and how do they impact the final performance?

\emph{RQ2}: Does our recommendation framework outperform the previous state-of-the-art recommendation models in terms of predicting implicit feedback?

\emph{RQ3}: Can our proposed Gaussian embeddings well capture the preferences of the users with uncertainty, further resulting in better recommendation performance?

\subsection{Experimental Settings}
\subsubsection{Dataset Description}
We evaluated our models on two public benchmark datasets: MovieLens 1M (ml-1m)\footnote{https://grouplens.org/datasets/movielens/},and Amazon music (Music)\footnote{http://jmcauley.ucsd.edu/data/amazon/}. The detailed statistics of the two datasets are summarized in Table \ref{tab:stat}. In ml-1m dataset, each user has at least 20 ratings. In Music dataset, we only reserved the users who have at least 1 rating record given its sparsity. %We only use user/item ID as the input of our model.

\begin{table}[t]
%\vspace{-0.3cm}
    \centering\small
%\vspace{-0.2cm}
    \begin{tabular}{c c c c c}
    \toprule
     Dataset   & \# user & \# item & \# interaction & Sparsity \\
    \midrule
       ml-1m  &  6040 & 3706 & 1000209 & 0.9553 \\
       Music & 1776	& 12929	& 46087 & 0.9980 \\  
    \bottomrule
    \end{tabular}
   \vspace{-0.2cm}
     \caption{Statistics of experimental datasets.}   \label{tab:stat}
\vspace{-0.2cm}
\end{table}

\subsubsection{Evaluation Protocols}
Following \cite{NAIS,NCF}, we adopted the \textit{leave-one-out evaluation}. We held out the latest one interaction of each user as the positive sample in test set, and paired it with 99 items randomly sampled from unobserved interactions. For each positive sample of every user in training set, we randomly sampled 4 negative samples. We then predicted and evaluated the 100 user-item interactions of each user in test set. We used two popular metrics evaluation measures, i.e., Hit Ratio (HR) and Normalized Discounted Cumulative Gain (nDCG) \cite{nDCG} to evaluate the recommendation performance of all compared models. The ranked list is truncated at 3 and 10 for both measures. %HR@k measures whether the testing item is on the top-k list or not. In addition, $nDCG=\frac{1}{Z}\sum_{i=1}^n\frac{2^{rel(i)}-1}{\log_2(i+1)}$ where Z is a normalized factor. 
Compared with Hit Ratio, nDCG is more sensitive to rank position because it assigns higher scores for top position ranking.
\subsubsection{Baselines}
%We compared our framework with the following state-of-the-art recommendation models in the experiments.
1. \emph{MF-BPR}: This model optimizes the standard MF with the pairwise Bayesian Personalized Ranking (BPR for short) loss \cite{BPR}.

2. \emph{NCF}: This model \cite{NCF} has been proven to be a powerful DNN-based CF framework consisting of a GMF (generalized matrix factorization) layer and an MLP (multi-layer perceptron). Both GMF and MLP are fed with user and item representations initialized in random. NCF parameters are learned based on obtained user-item interactions.

3. \emph{ConvNCF}: This is an improved version \cite{he2018outer} of NCF which uses outer product to explicitly model the pairwise correlations between the dimensions of the fixed point embedding space, and then applies multi-layer CNN to extract signal from the interaction map.

4. \emph{DeepCF}: This is a deep version \cite{DCF} of CF, aiming to fuse representation learning based methods and matching function based methods. It employs MLP to learn the complex matching function and low-rank relations between users and items.

5. \emph{NAIS}: In this framework \cite{NAIS}, a user's representation is the attentive sum of his/her historical favorite items' embeddings. A historical item's attention is computed based on the similarity between it and the candidate item. Thus such representations especially for the users with many historical favorite items, are also flexible w.r.t. different candidate items. %We compared our framework with this baseline to justify Gaussian embeddings are more adaptive to the users with uncertainty.

6. \emph{GER}: To the best of our knowledge, this baseline \cite{dos2017gaussian} is the only Gaussian embedding based recommendation model. It replaces dot product of vectors by inner product between two Gaussian distributions based on BPR framework. As we stated before, such ranking-based loss is not to applicable to other recommendation tasks.

7. \emph{MoG}: This is a variant of the model in \cite{joon2019iclr}, which averages predefined \textit{soft contrastive loss} between vector pairs to obtain matching probability between stochastic embeddings. We set its stochastic mappings to Gaussian embeddings. We compared MoG with our framework to highlight the effectiveness of computing matching probability based on convolutional operations.

In addition, we denote our framework as \emph{GeRec}. In order to achieve a fair comparison, we set the embedding dimension $D$=64 in all above baselines. The code package of implementing our framework is published on \\\emph{\url{https://github.com/JunyangJiang/gaussian-recommender}}.
%Moreover, we train the models with randomly sampled 4 negative items in all experiments and pre-train embedding layers of NCF, ConvNCF and DeepCF using the same MF-BPR loss.

\subsection{Experimental Results}
%In this subsection, we display evaluation results about our framework and related compared models, based on which we make some discussions.

\subsubsection{Hyper-parameter Tuning}
At first, we try to answer RQ1 through the empirical studies of hyper-parameter tuning in our framework. Due to space limitation, we only display the results of tuning three critical hyper-parameters of our framework GeRec, i.e., embedding dimension $D$, Monte-Carlo sampling number $K$ and our CNN's kernel number $T$, which were obtained from the evaluation on MovieLens dataset. Compared with previous deep models, only $K$ is additionally imported into our framework. Please note that when we tuned one hyper-parameter, we set the rest hyper-parameters to their best values. Table \ref{tab:HPres} displays our framework's performance of movie recommendation under different hyper-parameter settings. In general, larger $D$ and $T$ result in better performance. But we only selected $D=64$ and $T=64$ in our experiments given model training cost. And we set $K=9$ in the following comparison experiments according to the results in Table \ref{tab:HPres}. In addition, $D'$ is also set to 64. 
\begin{table}[t]
%\vspace{-0.2cm}
\centering\small
\begin{tabu}
{|p{20pt}|p{20pt}|p{40pt}|p{42pt} |}
\tabucline[1.2pt]{-}
Para. & Value & HR@10 & nDCG@10 \\
%\hline
%\multicolumn{4}{|c|}{\textbf{KGEBD}}\\
\tabucline[1pt]{-}
\multirow{4}*{$D$}&8& 0.7300& 0.4531\\
\cline{2-4} 
&16& 0.7311& 0.4627\\
\cline{2-4} 
&32& 0.7389& 0.4687\\
\cline{2-4} 
&\textbf{64}& \textbf{0.7474}& \textbf{0.4807} \\
\tabucline[1pt]{-}
\multirow{4}*{$K$}&1& 0.7235& 0.4459 \\
\cline{2-4}
&5& 0.7315& 0.4624\\
\cline{2-4} 
&\textbf{9}& \textbf{0.7474}& \textbf{0.4807} \\
\cline{2-4}
&13& 0.7452& 0.4799 \\
\tabucline[1pt]{-}
\multirow{4}*{$T$} & 8& 0.7278& 0.4624 \\
\cline{2-4} 
&16& 0.7310& 0.4629 \\
\cline{2-4} 
&32& 0.7370& 0.4679 \\
\cline{2-4} 
&\textbf{64}& \textbf{0.7474}& \textbf{0.4807}\\
\tabucline[1.2pt]{-}
\end{tabu}
\vspace{-0.2cm}
\caption{{GeRec's hyper-parameter tuning results on MovieLens.}}\label{tab:HPres}
\vspace{-0.3cm}
\end{table}

\subsubsection{Global Performance Comparisons}
To answer RQ2, we compared our framework with the baselines in terms of recommendation performance. The results listed in Table \ref{tab:res} were the average scores of 5 runs, showing that our framework GeRec performs best on the two datasets. Specifically, GeRec's advantage over MF-BPR, NCF, ConvNCF and DeepCF shows that Gaussian embeddings represent users and items better than the embeddings of fixed points, resulting in more precise recommendation results. GeRec's advantage over NAIS shows that although attention-based user representations are also flexible embeddings, they do not perform as well as Gaussian embeddings. GeRec's superiority over GER and MoG justifies that, our CNN-based evaluation of the correlations between the Gaussian samples of users and items is more effective than the operations in GER and MoG.

\begin{table*}[t]
%\vspace{-0.2cm}
\small
\begin{center}
%\vspace{-0.5cm}
\begin{tabular}
{|p{40pt}|p{35pt}|p{37pt}|p{35pt} |p{42pt}|p{35pt}|p{37pt}|p{35pt}|p{42pt}|}
\hline
\multirow{2}{*}{\textbf{Model}}& \multicolumn{4}{c|}{MovieLens 1M} &\multicolumn{4}{c|}{Amazon Music}\\
\cline{2-9} & HR@3 & nDCG@3 & HR@10& nDCG@10 & HR@3 & nDCG@3 & HR@10& nDCG@10 \\
\hline
MF-BPR& 0.3996 &  0.3085& 0.6760 & 0.3978& 0.1536 & 0.1198 & 0.2711 & 0.1448\\
\hline
NCF& 0.4739& 0.3685& 0.7288& 0.4652& 0.1777& 0.1336& 0.3358& 0.1913\\
\hline
ConvNCF & 0.4772 & 0.3622 & 0.7290 & 0.4597 & 0.1758 & \textbf{0.1431} & 0.3370 & 0.1990 \\
\hline
DeepCF & 0.4755&  0.3823 & 0.7326& 0.4680& 0.1798& 0.1396& 0.3416& 0.1952\\
\hline
NAIS& 0.4497 & 0.3618 & 0.7182 & 0.4418 & 0.1703 & 0.1317 &0.2852 &0.1721 \\
\hline
GER & 0.4016 & 0.3171 & 0.7018 & 0.4264 & 0.1541 & 0.1258 & 0.2953 & 0.1489 \\
\hline
MoG& 0.4586& 0.3669& 0.7245& 0.4625& 0.1716& 0.1309& 0.3196& 0.1791\\
\hline
\textbf{GeRec}& \textbf{0.4841}& \textbf{0.3846}& \textbf{0.7474}& \textbf{0.4807}& \textbf{0.1863}& {0.1429}& \textbf{0.3464}& \textbf{0.2034}\\
\hline
\end{tabular}
\vspace{-0.2cm}
\caption{Global Recommendation Performance results show that our GeRec outperforms all baselines on the two datasets.}\label{tab:res}
\end{center}
\vspace{-0.2cm}
\end{table*}

\subsubsection{Effectiveness on Capturing User Uncertainty}
To answer RQ3, we evaluated our framework particularly against the users with uncertain preferences. At first, we introduce how to categorize such users. As stated in Sec. 1, we focus on two kinds of users with uncertainty in this paper. The first kind of such users are those with sparse observed user-item interactions, because very little information about their preferences can be obtained from their historical actions. The second kind of such users are those having many distinct preferences, because we can not identify which genre of preference is their most favorite one. 
\begin{table}[htbp]
\vspace{-0.1cm}
\small
\begin{center}
%\vspace{-0.4cm}
\begin{tabular}
{|p{40pt}|p{40pt}|p{40pt}|p{40pt} |}
\hline
\multicolumn{2}{|c|}{1st kind of uncertain users} &\multicolumn{2}{c|}{2nd kind of uncertain users}\\
\hline
 $o_1$ &  variance &  $o_2$ &  variance \\
 \hline
1.1$\sim$ 1.5& 0.790 & 0.9$\sim$1& 1.057 \\
\hline
1.5$\sim$ 1.9& 0.796 & 0.8$\sim$ 0.9& 1.003 \\
\hline
1.9$\sim$ 2.3& 0.778 & 0.7$\sim$ 0.8& 0.855 \\
\hline 
2.3$\sim$ 2.7& 0.746 & 0.6$\sim$ 0.7& 0.801 \\
\hline
2.7$\sim$ 3.1& 0.549 & 0.5$\sim$ 0.6& 0.754 \\
\hline
3.1$\sim$ 3.5& 0.435 & 0.4$\sim$ 0.5& 0.754 \\
\hline
\end{tabular}
\vspace{-0.2cm}
\caption{The learned Gaussian variances for MovieLens users. 
%show that user uncertainty is captured by Gaussian embeddings precisely.
}\label{tab:Unres}
\end{center}
\vspace{-0.6cm}
\end{table}

Inspired by \cite{zhu2018deep}, we identified these two kinds of uncertain users according to two metrics, respectively. Specifically, for the first kind of users, we filtered out six user groups according to a metric $o_1$. The users of $o_1$ are those who have $10^{o_1}$ observed user-item interactions. Thus small $o_1$ indicates the users with more the first kind of uncertainty. For the second kind of users, we also filtered out six user groups according to metric $o_2$. We compute $o_2$ for a given user $u$ as follows. For each pair $(m_i,m_j)$ of movies rated by $u$, suppose $G_i$ and $G_j$ are the genre sets of $m_i$ and $m_j$, respectively. Then, we set
$
o_{ij}=1-\frac{|G_i\cap G_j|}{|G_i\cup G_j|}
$. Finally, we use average $o_{ij}$ of all movie pairs as $u$'s $o_2$. As a result, large $o_2$ indicates more preference diversity, i.e., the second kind of uncertainty. For space limitation, we only display the results of MovieLens users in Table \ref{tab:Unres}. In the table, the displayed variances are the average Gaussian variances learned by our framework, showing that our proposed Gaussian embeddings assign larger variances to the users with more uncertainty. Thus, such distribution based embeddings represent the users with uncertainty well, resulting in better recommendation performance.

%\vspace{-0.1cm}
\section{Related Work}
%\vspace{-0.1cm}
%In this section, we introduce some works related to our work.
%\paragraph{Matrix factorization based recommendation:}
MF-based models constitute one important family of recommendation models, such as latent factor model \cite{MF} and non-negative matrix factorization \cite{NFM}. Based on these traditional MF-based models, some improved versions had been proposed and proven more effective. For example, SVD++ \cite{SVD++} improves SVD through taking into account the latent preferences of users besides explicit user-item interactions. MF-BPR optimizes standard MF with pairwise Bayesian Personalized Ranking \cite{BPR} loss. Factorization Machine (FM) \cite{FM} captures the interactions between user/item features to improve performance of model learning. All these models represent users and items by a vector containing the latent factors of users/items, of which the values are fixed once they are learned from user-item interactions, so are not adaptive to the users/items with uncertainty. 

%\vspace{-0.1cm}
%\paragraph{Deep learning based recommendation:} 
In recent years, many researchers have justified that traditional recommendation models including CF and MF-based models, can be improved by employing DNNs. For example, the authors in \cite{AutoRec} proposed a novel AutoEncoder (AE) framework for CF. DeepMF \cite{DMF} is a deep version of MF-based recommendation model. In addition, NCF model \cite{NCF} integrates generalized matrix factorization model (GMF) and multiple-layer perceptron (MLP) to predict CF-based implicit feedback. DeepCF \cite{DCF} also employs MLP to learn the complex matching function and low-rank relations between users and items, to enhance the performance of CF. In general, these deep models also represent users/items by embeddings which are used to feed the neural networks, and their embeddings also correspond to fixed points in embedding space without flexibility. Although the models in\cite{DRM,NAIS} import attention mechanism to make user representations more flexible, such attention-based embeddings were proven not so good as Gaussian embeddings by our experiments.

%\vspace{-0.1cm}
%\paragraph{Gaussian embedding:}
Gaussian embeddings are generally trained with ranking objective and energy functions, such as probability product kernel and KL-divergence. The authors in \cite{vilnis2014word} first used a max-margin loss to learn word representations in the space of Gaussian distributions to model uncertainty. Similarly, \cite{he2015learning} and \cite{dos2016multilabel} learn Gaussian embeddings for knowledge graphs and heterogeneous graphs, respectively; \cite{dos2017gaussian} uses Gaussian distributions to represent users and items in ranking-based recommendation. To improve graph embedding quality, \cite{bojchevski2018deep} takes into account node attributes and employs a personalized ranking formulation, and \cite{zhu2018deep} incorporates 2-Wasserstein distance and Wasserstein Auto-Encoders. All these methods employ ranking function and thus can not be applied to other recommendation tasks easily. Recently, \cite{joon2019iclr} learns stochastic mappings of images with contrastive loss and also uses Gaussian embeddings.

%\vspace{-0.1cm}
\section{Conclusion}
%\vspace{-0.1cm}
In this paper, we propose a unified recommendation framework in which each user or item is represented by a Gaussian embedding instead of a vector corresponding to a single fixed point in feature space. Moreover, convolutional operations are adopted to effectively evaluate the correlations between users and items, based on which precise recommendation results of both personalized ranking and rating prediction can be obtained. Our extensive experiments not only demonstrate our framework's superiority over the state-of-the-art recommendation models, but also justify that our proposed Gaussian embeddings capture the preferences of the users with uncertainty very well.

%\newpage
\bibliographystyle{named}
\bibliography{refer}
\end{document}